\documentclass[aps,pre,showkeys,preprint]{revtex4}
\usepackage{graphicx}
\usepackage{amsmath}
\usepackage{amsfonts}
\usepackage{amssymb}
\usepackage{bm}
\usepackage{enumitem}  
\usepackage{float}
\usepackage{xcolor}
\usepackage{color}

\begin{document}


\title{Dynamics of antiproton plasma in a time-dependent harmonic trap}

\author{L. G. F. Soares and F. Haas}
\affiliation{Instituto de F\'{\i}sica, Universidade Federal do Rio Grande do Sul, Av. Bento Gon\c{c}alves 9500, 91501-970 Porto Alegre, RS, Brasil.}


\begin{abstract}
An antiproton plasma confined in a quasi-1D device is described  in terms of a self-consistent fluid formulation using a variational approach. Unlike previous treatments, the use of the time-dependent variational method allows to retain the thermal and Coulomb effects. A certain {\it Ansatz} is proposed for the number density and fluid velocity fields, which reduces the problem essentially to ordinary nonlinear differential equations.  In    adiabatic  cooling,  the  frequency  of  the  trap  potential  is  slowly decreased.   An  adiabatic  equation  of  state is assumed for closure. The numerical simulation of the nonlinear dynamics is performed, for realistic parameters. 

\end{abstract}


\maketitle

One of the goals of modern physics is the creation of large samples of cold antihydrogen (composed of an antiproton and a positron) for studies on CPT invariance and gravitational influence on antimatter \cite{Gabrielse1}. For this, the confinement and cooling of antiprotons is an important step  \cite{Gabrielse2} and the adiabatic cooling technique leads \cite{Gabrielse3} to temperatures of 3 K. In this Brief Communication,  a non-relativistic antiproton plasma confined in a one-dimensional trap is studied when  thermal and Coulomb effects are both relevant. For this purpose, a fluid formulation and a time-dependent variational method are applied, allowing us to assess the  time-dependent dynamics by adopting a Gaussian {\it Ansatz}. The limit cases, where the thermal or Coulomb interactions can be neglected were explored in {Ref. [4], which uses a non-variational approach. The Lagrangian variables with a linear velocity hypothesis in the fluid frame lead to a local or a constant number density distribution, which restrain the results to the limit cases \cite{gus2}. By adopting a Lagrangian density with a Gaussian {\it Ansatz} this restriction is removed.}

The confinement of single electrically charged particles is made by  cylindrical devices called Penning-Malmberg traps. In these traps, the quadrupolar electrostatic field (originated by hyperbolic electrodes) creates a  potential well that confines the antiproton plasma in the axial direction, and a homogeneous magnetic field confines it radially \cite{Malmberg,Gould}. 

The cooling of electrically charged particles is achieved by several methods and techniques \cite{Rolston}. After injected in the trap, the collisional cooling is used for pre-cooling the antiprotons and is done with positrons \cite{Gabrielse3} or electrons \cite{Manfredi}. This cooling process is based on the thermal equilibrium reached after charged particles (with different temperatures) collide with each other. After that,  evaporative cooling or  adiabatic cooling is required to obtain  temperatures of the order of a few Kelvins. 

In evaporative cooling, the elastic collisions scatter highly energetic particles out of the potential well leading to a lower antiproton density \cite{Andresen}. On the other hand, in the adiabatic cooling, the external harmonic confinement  has a slowly decreasing frequency. For that reason, the restoring force makes the plasma expand adiabatically, and the temperature decreases. In this process, almost no losses are observed \cite{Manfredi,Gabrielse3}. Moreover, the adiabatic cooling was also applied in atom cooling by lowering the standing-wave intensity \cite{Chen}, in optical lattice by lowering the lattice light intensity \cite{Kastberg} and in  electrically trapped polar molecules \cite{Englert}.

The study of one-component plasma dynamics is a traditional field \cite{Davidson}. The  analysis of exact or approximate nonlinear structures can be simplified using variational methods, as in Bose-Einstein condensates \cite{fer2,Salasnich,Salasnich2,Perez,Perez2,Ad,Gh} and quantum electron gases \cite{fer1,fer3,fer4,fer5}.  In this context, our treatment consists of the minimization of the action functional, reducing the problem to a set of coupled ordinary differential equations by adopting a trial function. In the hydrodynamic model, the external 
time-dependent potential provides confinement, while thermal and Coulomb effects tend to expand the gas. {In the quasi-1D model, collisional effects are not included in the model equations, since a very collisional plasma would not keep its quasi-1D character for a long time.}  Thermal effects are also considered, for the sake of generality.

The hydrodynamic equations for the antiproton plasma trapped in an one-dimensional time-dependent well are
\begin{eqnarray}\label{1.a}
\frac{\partial n}{\partial t} + \frac{\partial}{\partial z}(nv) &=& 0\,, \\
\label{1.b}
\frac{\partial v}{\partial t} + v\frac{\partial v}{\partial z} &=& - \frac{1}{m n}\frac{\partial p}{\partial z} -\frac{1}{m}\frac{\partial }{\partial z}V_c+ \frac{e}{ m}\frac{\partial \phi}{\partial z}\,,\\
\label{1.c}
\frac{\partial^2 \phi }{\partial z^2} &=& \frac{e \sigma_\perp n}{\varepsilon_0}\,.
\end{eqnarray}
The system is composed of antiprotons (mass $m$ and charge $-e$)  with a 1D number density $n(z,t)$, a fluid velocity $v(z,t)$ and a self-consistent electrostatic potential $\phi(z,t)$, where $\sigma_\perp$ is the 2D number density in the perpendicular plane and $\varepsilon_0$ is the vacuum permittivity. Moreover, $V_c$ is a time-dependent confining potential. Since the motivation of this paper is about the adiabatic cooling of antiprotons,  a decreasing time-dependent harmonic potential will be adopted in the form
\begin{equation}\label{1.c0}
V_c =\frac{m\omega^2(t)z^2}{2},\,\,\,\,\omega(t)=\frac{\omega_0}{(1 + \Omega t)^{\beta}} \,,
\end{equation}
where $\omega_0, \Omega$ and $\beta$ are positive constants. For slowly varying angular frequency one has $|\dot\omega|/\omega \ll \omega$, or $\beta\Omega \ll \omega_0 (1 + \Omega t)^{1-\beta}$, which holds \cite{Manfredi} for all times $t \geq 0$ provided $\beta \leq 1$ and $\beta\Omega \ll \omega_0$. Under these conditions, the energy of the system remains approximately constant in a period of an appropriate time scale. Additionally, the pressure $p(z,t)$ must be related to the antiprotons  density to closure the set of fluid equations. Since the energy barely changes in time, an adiabatic equation of state is assumed. In the present case   
\begin{equation}
p = n_0 \kappa_B T_0 \left(\frac{n}{n_0}\right)^3
\end{equation}
is the isentropic equation of state with adiabatic index $\gamma = (d+2)/d = 3$ for the dimensionality $d = 1$, where $n_0$ reference number density (that will be better defined later) and $T_0$ is a  reference temperature ($\kappa_B$ is the Boltzmann constant). {The number density $n$ is assumed to satisfy decaying boundary conditions, in view of the confinement. The velocity field and the scalar potential have more free boundary conditions, as long as they are consistently determined from the continuity and Poisson equations, given the number density.} 

The problem of solving the set of equations (\ref{1.a})-(\ref{1.c})   can be reinterpreted as a variational problem corresponding to the minimization of the action functional $S=\int dt\, {dz} \,\mathcal{L}$,  specified by  the Lagrangian density 
\begin{eqnarray}\label{1.d}
\mathcal{L}&=&mn\bigg[\frac{1}{2}\bigg(\frac{\partial \theta}{\partial z} \bigg)^2+\frac{\partial \theta}{\partial t}\bigg]+n(V_c-e\phi)-\frac{\varepsilon_0}{2\sigma_\perp}\bigg(\frac{\partial \phi}{\partial z}\bigg)^2\nonumber\\
&+&\int dn \int \frac{d p}{n}\,,
\end{eqnarray}
where the independent fields are  the velocity potential $\theta = \theta({z},t)$, so that ${v}=\partial\theta/\partial z$, the number density $n$, 
and the self-consistent electrostatic
 potential  $\phi$. Indeed, it can be easily shown that the minimization of the respective Lagrangian density corresponding to Eq. (\ref{1.d}), with respect to the fields $\theta$, $n$ and $\phi$ respectively yields the continuity, momentum and Poisson equations. 

A normalized Gaussian {\it Ansatz} is adopted, 
\begin{eqnarray}\label{1.e}
    n(z,t)=\frac{A}{\alpha}\exp\bigg({-\frac{\rho^2}{2}}\bigg)\,,
\end{eqnarray}
where the constant $A=N/(\sqrt{2\pi}A_\perp \sigma_\perp)$ is related to the total number of trapped antiprotons, $N=A_\perp \sigma_\perp\int_{-\infty}^\infty n(z,t)dz$ is the number of confined antiprotons, {$A_\perp$ is the occupied area in the perpendicular plane}   and 
\begin{equation}\label{1.e.1}
    \rho = \rho(z,t)= {\frac{z-d(t)}{\alpha(t)}}\,.
\end{equation}
The Gaussian form reflects the plasma confinement and is amenable for analytic treatment. 
The time-dependent  coordinates $d(t)$ and $\alpha(t)$, respectively give the position of the center of mass (dipole) and the width of the atomic cloud in the $z$ direction. 
In addition,  define the reference number density as $n_0={N}/({\sqrt{2\pi}\alpha_{0}A_\perp\sigma_\perp})$, where $\alpha_{0}=\alpha(0)$.  

Direct substitution of the {\it Ansatz} in the continuity equation leads to an exact solution for the velocity field, given by  
\begin{equation}\label{1.f}
    v= \frac{\dot{\alpha}}{\alpha}(z-d)+\dot{d} \,,
\end{equation}
{ignoring for simplicity an additive arbitrary function of time only.} 
 Since $v=\partial\theta/\partial z$, the  {velocity potential} $\theta$ in the Lagrangian density can be written as
\begin{equation}\label{1.g}
\theta=\frac{\dot{\alpha}}{2\alpha}(z-d)^2+\dot{d}(z-d)\,,
\end{equation}
where an extra {irrelevant gauge function of time only} was ignored.

In addition, by direct integration of Eq. (\ref{1.c}), yields 
\begin{equation}\label{1.h}
   \phi=\frac{N e }{2A_\perp\varepsilon_0}\bigg[\sqrt{\frac{2}{\pi}}\alpha e^{-\frac{\rho^2}{2}} +(z-d){{ {\rm Erf}}\bigg(\frac{\rho}{\sqrt{2}}\bigg)}\bigg] + {c1(t)z + c2(t)},
\end{equation}
where ${\rm Erf}(s)=(2/\sqrt{\pi})\int_0^s e^{-s'^2}ds'$ denotes the error function of a generic argument $s$. {To determine the functions $c_{1,2}(t)$ we follow the trend of Ref. [25] choosing $\phi(\pm D)=0$ where $D$ is the size of the system, eventually set to infinity at the end of the calculation. This yields $c_1(t)=0$ and allows to write 
\begin{equation}
\label{bc}
\lim_{D\rightarrow\infty}\int_{-D}^{D} dz\,\left(e n \phi + \frac{\varepsilon_0}{2\sigma_\perp}\left(\frac{\partial\phi}{\partial z}\right)^2\right) = \frac{e}{2}\lim_{D\rightarrow\infty}\int_{-D}^{D}dz\, n\, \phi \,,
\end{equation}
with a vanishing surface term and assuming Poisson's equation to be valid.} 
Therefore, the electric field is given by  
\begin{equation}
    {E = - \frac{\partial\phi}{\partial z} = -\frac{N e}{2A_\perp\varepsilon_0}{{\rm Erf}\bigg(\frac{\rho}{\sqrt{2}}\bigg)} \,.}
\end{equation}

In order to derive the dynamical behavior of the new coordinates, the Lagrangian is computed. After the substitution of  Eqs. (\ref{1.e}), (\ref{1.g}) and (\ref{1.h}) into Eq. (\ref{1.d}), one has 
\begin{eqnarray}\label{1.l0}
    L(d,\dot{d},\alpha,\dot{\alpha})&=&
    -\frac{A_\perp\sigma_\perp}{m N}\int\mathcal{L} {dz}\nonumber\\
    &=&\frac{1}{2}(\dot{d}^2 + \dot{\alpha}^2)-U_d-U_\alpha
\end{eqnarray}
where 
\begin{equation}
    U_d=\frac{\omega^2(t)}{2}d^2
\end{equation}
and
\begin{equation}\label{1.l}
    U_\alpha=\frac{\omega^2(t)}{2}\alpha^2+\frac{a}{2}\frac{1}{\alpha^2}-b\alpha\,.
\end{equation}
 which are, respectively, the pseudo-potentials corresponding to the dipole  and oscillating width modes,  where the constants $a={\kappa_B{T_0}N^2}/({2\sqrt{3}\pi mn_0^2A_\perp^2\sigma_\perp^2})$ and $b={Ne^2}/(2\sqrt{\pi}m\varepsilon_0A_\perp)$ are introduced. The Lagrangian in Eq. (\ref{1.l0}) only depends on two degrees of freedom, namely the dipole and the width. Also, in Eq. (\ref{1.l}),
the term $\sim \omega^2(t)$ is related to the time-dependent harmonic confinement, the term $\sim b$ corresponds to the self-consistent electrostatic potential and the term $\sim a$ is due to the adiabatic pressure. {To evaluate the electrostatic part of the Lagrangian, Eq. (\ref{bc}) was used. The resulting integral is divergent, but with a divergence not depending on the dynamical variables $d, \alpha$ and therefore ignorable \cite{fer5}. Moreover a total time derivative term was discarded.}

Since the Lagrangian is obtained, one can apply the Euler-Lagrange equations for each variational parameter thus deriving the  equations of motion. The dynamics of the center of mass is given by 
\begin{equation}\label{2.a0}
  \ddot{d}+\omega^2(t)d=0\,,  
\end{equation}
  which, as  can directly be seen, is decoupled to the width equations showing time-dependent oscillations around the origin. Furthermore, this motion is linear and independent of the number of atoms. The solution of this equation  can be mapped  in terms of Bessel functions for $0<\beta<1$ (when Eq. (\ref{1.c0}) is valid) or in terms of approximate WKB (Wentzel-Kramers-Brillouin) solutions for slowly varying frequency  \cite{gus}. 
  
The equation of motion for the oscillating width in normalized variables $\bar{\alpha}=\alpha/\alpha_0$ and $\tau=\omega_0t$ is
\begin{equation}\label{2.a1}
    \ddot{\bar{\alpha}}+\bar{\omega}^2(\tau)\bar{\alpha}=\frac{1}{\omega_0^2}\bigg(\frac{\omega_{T}^2}{\bar{\alpha}^3}+\frac{\omega_p^2}{\sqrt{2}}\bigg)\,,
\end{equation}
or
\begin{equation}\label{2.a}
    \ddot{\bar{\alpha}}=-\frac{d U}{d\bar{\alpha}}\,,
\end{equation}
where $\bar{\omega}(\tau)=(1+\Omega/\omega_0 \,\tau)^{-\beta}$ and $U$ is the pseudo-potential defined by 
\begin{equation}\label{2.b}
    U=\frac{\bar{\omega}^2(\tau)}{2}\bar{\alpha}^2+\frac{1}{\omega_0^2}\bigg(\frac{\omega_T^2}{2}\frac{1}{\bar{\alpha}^2}-\frac{\omega_p^2}{\sqrt{2}}\bar{\alpha}\bigg)\,,
\end{equation}
where $\omega_T=\sqrt{\kappa_BT_0/(\sqrt{3}m\alpha_0^2)}$ and  $\omega_p=\sqrt{n_0\sigma_\perp e^2/(m\varepsilon_0)}$ is the plasma frequency, { which are respectively related to thermal and self-consistent (Coulomb) effects.  }  

The confining potential $V_c$ manifests itself in the
harmonic forces on the left-hand side of both Eqs. (\ref{2.a0}) and
(\ref{2.a1}). As expected, the equations for $d$ and $\alpha$ decouple for
purely harmonic confinement. From Eq. (\ref{2.a1}), the  oscillating width is described by a forced Pinney equation. { Similar equations were obtained for a many-electron dynamics in  a semiconductor quantum well \cite{fer5} using a self-consistent quantum hydrodynamic model (QHM) and for a trapped  non-neutral
plasma  \cite{sh}, except that Eq. (\ref{2.a1}) has a time-dependent confining potential.}  The non-linearity comes from the repulsive interactions due to the pressure and self-consistent interaction terms.  
From the shape of the pseudo-potential (Fig.$\,\,1$), one has
that $\alpha$ will always execute time-dependent  oscillations around the minimum that grows  as the frequency decreases in time. 

Equation (\ref{2.a}) can be numerically solved for realistic and accessible antiproton plasmas parameters \cite{Enomoto, Gabri}, namely: $\kappa_bT_0 = 30\,\mathrm{eV}$ and $\alpha_0= 5\,\mathrm{cm}$. Supposing $N=10^5$ confined antiprotons with
a circular cross section of radius $2\,\mathrm{mm}$, yields a number density $n_0\sigma_\perp=6.3\,\times10^{10}\,\mathrm{m^{-3}}$ so that $\omega_T/2\pi=1.5\,\times10^5\,\mathrm{Hz}$ and $\omega_p/2\pi= {5.3\,\times10^4}\, \mathrm{Hz}$. In addition, the trap frequency, Eq. (\ref{1.c0}), is considered, with $\beta=1$, $\Omega=0.02\omega_0$ and $\omega_0/2\pi=100 \,\mathrm{KHz} $  . The resulting  nonlinear oscillation is shown in Fig. $2$. The  oscillation amplitude grows in time showing the expansion of the plasma as the frequency is slowly decreased in time.

 \begin{figure}[ht]
\label{fig1}
\begin{center}
\includegraphics[width=4.0in]{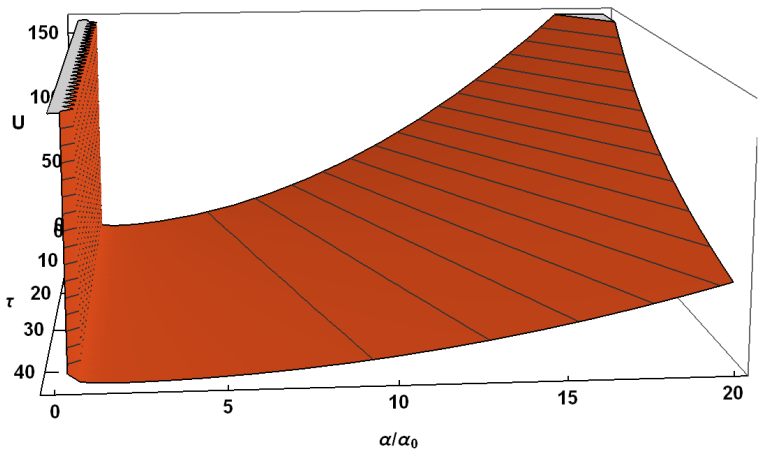}
\caption{Pseudo-potential from Eq. (\ref{2.b}). Parameters are indicated in the text.}
\end{center}
\end{figure}

 \begin{figure}[ht]
\label{fig2}
\begin{center}
\includegraphics[width=4.0in]{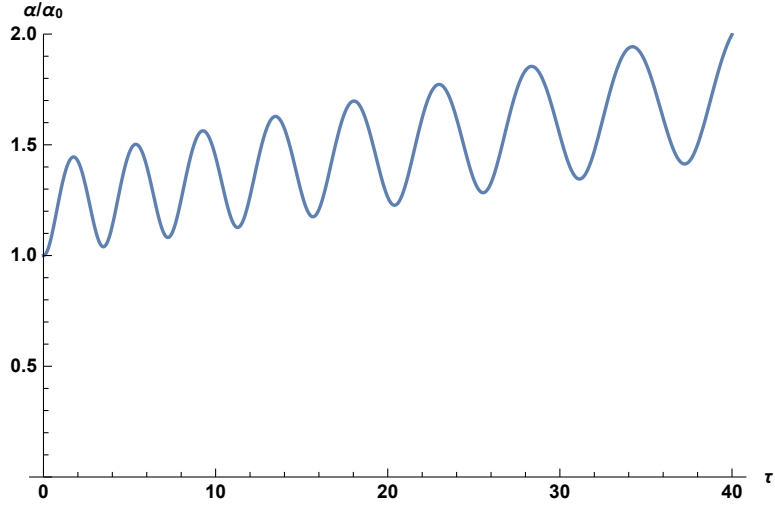}
\caption{Numerical solution of Eq. (\ref{2.a}). Parameters are indicated in the text.  Initial conditions: $\alpha=1$ and  $\dot{\alpha}_{0}=0$.}
\end{center}
\end{figure}

In this Brief Communication, confined  antiproton plasmas in a quasi-1D geometry have been studied. The main result is the  nonlinear analysis of an  antiproton plasmas in a time-dependent trap  with a slowly decreasing
frequency. This trap provides the adiabatic cooling of the trapped non-neutral plasma. When thermal and Coulomb effects are both relevant, the basic equations for the quasi-1D  variational description under arbitrary confinement 
is reducible to a forced Pinney equation. For this purpose, the starting point was a hydrodynamic set of equations reinterpreted in terms of the minimization of an action functional, adopting a Gaussian {\it Ansatz}. The time-dependent variational method with an adiabatic equation of state retains both the thermal and electrostatic effects.     
 Moreover,  the results have been applied to nonlinear oscillations compatible with typical experiments. The results will be relevant for trapped non-neutral plasmas under time-varying harmonic potentials and are also relevant for the experimental creation of antihydrogen atoms. In addition,  the present approach can be directly adapted to damped non-neutral  confined  plasmas,  where  the  damping  mechanism  can  be  traced  back  to collisions with neutrals or  to adiabatic cooling of neutral atoms by lowering the current intensity in the  anti-Helmholtz coils.

\section*{Acknowledgments}
L.~G.~F.~S.~ and F.~H.~acknowledge the support of the Con\-se\-lho Na\-cio\-nal de De\-sen\-vol\-vi\-men\-to Cien\-t\'{\i}\-fi\-co e Tec\-no\-l\'o\-gi\-co (CNPq). This study was financed in part by the Coordena\c{c}\~ao de Aperfei\c{c}oamento de Pessoal de N\'{\i}vel Superior - Brasil (CAPES) - Finance Code 001. 

\section*{DATA AVAILABILITY}

The data that support the findings of this study are available from the corresponding author upon reasonable request.

\end{document}